\documentclass[prd,amsmath,amssymb,showpacs]{revtex4}
\usepackage{graphicx}
\usepackage{color}
\usepackage[colorlinks=true,linkcolor=red]{hyperref}
\usepackage{hyperref}

\begin{document}

\hfill{ITEP-TH-19/06}

%\vspace{5mm}

\title{Thermal radiation of various gravitational
backgrounds}
\author{Emil T.Akhmedov}
\email{akhmedov@itep.ru}
\affiliation{Moscow, B.Cheremushkinskaya, 25, ITEP, Russia 117218}
\author{Valeria Akhmedova}
\email{lera@itep.ru} \affiliation{Physics Department, CSU Fresno, Fresno, CA 93740-8031}
\author{Terry Pilling}
\email{terry@droog.sdf-eu.org} \affiliation{Department of Physics, North Dakota State University,
Fargo, ND 58105}
\author{Douglas Singleton}
\email{dougs@csufresno.edu} \affiliation{Physics Department, CSU Fresno,
Fresno, CA 93740-8031}

\date{\today}

\begin{abstract}
We present a simple and general procedure for calculating the
thermal radiation coming from any stationary metric. The physical
picture is that the radiation arises as the quasi--classical
tunneling of particles through a gravitational barrier. We study
three cases in detail: the linear accelerating observer (Unruh
radiation), the non-rotating black hole (Hawking radiation), and
the rotating/orbiting observer (circular Unruh radiation). For the
linear accelerating observer we obtain a thermal spectrum with the
usual Unruh temperature. For the non-rotating black hole we obtain
a thermal spectrum, but with a temperature twice that given by the
original Hawking calculations. We discuss possible reasons for the
discrepancies in temperatures as given by the two different
methods. For the rotating/orbiting case the quasi--classical
tunneling approach indicates that there is no thermal radiation.
This result for the rotating/orbiting case has experimental
implications for the experimental detection of this effect via the
polarization of particles in storage rings.
\end{abstract}


\maketitle

%%%%%%%%%%%%%%%%%%%%%%%%%%%%%%%%%%%%%%%%%%%%%%%%%%%

\section{Introduction}

One of the most surprising results of doing quantum field theory
on a curved gravitational background was the theoretical discovery
that certain gravitational backgrounds gave rise to thermal
radiation from the vacuum. The first, prototypical example is the
Hawking radiation of a Schwarzschild black hole \cite{hawking}.
The original derivation given by Hawking is lengthy and complex
making it difficult to connect a concrete physical picture with
the calculation. Recently \cite{parikh1} \cite{parikh2} \cite{kraus}
\cite{vagenas} \cite{zerbini1} \cite{zerbini2} a simple derivation of Hawking
radiation was presented. In this note we present a similar
calculation based on the Hamilton-Jacobi equation. This
calculation is based on the physical picture of Hawking radiation
as a tunneling process -- particles tunnel out from behind the
horizon. The advantage of the present approach is that it can
easily be applied to any stationary metric. Using this procedure
we are able to make some interesting observations concerning the
Hawking temperature of Schwarzschild black hole. We find that the
temperature as calculated in the semi--classical tunneling
approach is twice that of the original Hawking calculations
\cite{hawking}. In addition we apply our procedure to the Rindler
\cite{unruh} and rotating/orbiting system \cite{letaw} and find
that the semi--classical tunneling approach indicates that one
does not observe thermal, Unruh-like radiation for circular
motion. This has experimental implications for the experimental
observation of this effect using the polarization of particles in
circular particle accelerators as ``thermometers" \cite{bell}. If
we take into account the considerations of \cite{AkhmSing} along with
the calculations of this note, we can conclude that the
quasi--classical tunneling approach gives correct answers only in
the case of stationary backgrounds with event horizons.

In this note we do not take into account the back--reaction of
gravity on the quantum fluctuations of all other fields. In this
context, to determine whether a particular gravitational background radiates
or not we solve the Klein--Gordon equation in a curved background:
\begin{equation}
\left[- \frac{\hbar^2}{\sqrt{-g}} \partial_\mu g^{\mu\nu} \sqrt{-g} \partial_\nu\,
+ m^2\right] \phi = 0. \label{KG}
\end{equation}
The signature of the metric is (-1,1,1,1)
and $ds^2 = g_{\mu\nu}(x) \, dx^\mu \, dx^{\nu}$, $g_{\mu\nu} \,
g^{\nu\alpha} = {\delta_{\mu}}^\alpha$.

We are looking for the solutions of \eqref{KG} having the form:
$\phi(x) \propto \exp\left\{-\frac{\rm i}{\hbar} \, S(x) +
\dots \right\}$.
Inserting this into \eqref{KG} and taking the limit $\hbar \to 0$ we find to
order $\hbar^0$ the following equation
\begin{equation}
g^{\mu\nu} \, \partial_\mu S \, \partial_\nu S + m^2 = 0,\label{HJ}
\end{equation}
which is just the relativistic Hamiltonian--Jacobi equation
for the classical action of a relativistic particle in the curved
background. The condition under which our approximation is
valid is worked out in \cite{landau}.

The metrics which we consider are stationary and, hence, have
time--like Killing vectors. We are going to look for the
particle--like solutions of \eqref{HJ} which behave as $S = E\, t
+ S_0(\vec{x})$ , where $x_\mu = (t,\vec{x})$. $E$ is the energy
of the particle. The wave function for such a solution behaves as
$\phi \propto e^{-\frac{\rm i}{\hbar} \, E\, t}$ and corresponds
to a state with definite energy. It is these states which are
supposed to be observed by detectors  \cite{unruh}.

If the solution $S_0(\vec{x})$ of \eqref{HJ} has a
non--zero imaginary part for some particle trajectory this implies
that the gravitational background in question is unstable with respect
to radiation of such a definite energy states.
What we show via our
procedure is that quantum mechanically
{\it if} there is a particle behind the horizon it can tunnel
through it with the given rate. It seems that if there are no particles behind
the horizon to begin with then we will not see any flux of
particles from behind the horizon. In quantum
field theory, however, the situation is different, because the number of
particles is not conserved. In general if the
corresponding calculation in the quantum mechanical limit shows a
tunneling, then in quantum field theory one sees a flux of
particles. In this case the wave function behaves as:
$\phi \propto e^{- \frac{1}{\hbar} \, {\rm Im} S_0}$,
which describes tunneling of the particle through the gravitational
barrier and leads to the decay rate of the background in question
as follows:
$\Gamma \propto |\phi|^2 \propto e^{-\frac{2}{\hbar} \, {\rm
Im} S_0 }$.
Now we are going to solve \eqref{HJ} for various well
known gravitational backgrounds.

\section{Schwarzschild black hole radiation}

We first look at the radiation coming from a Schwarzschild black hole with mass $M$.
Using the Schwarzschild background
\begin{equation}
ds^2 = - \left(1 - \frac{2M}{r}\right) \, dt^2 +
\frac{dr^2}{\left(1 - \frac{2M}{r} \right)} + r^2 d\Omega^2 , \label{schwarz}
\end{equation}
the Hamilton--Jacobi equation becomes
\begin{equation}
- \frac{1}{\left(1 - \frac{2M}{r}\right)}\,
\left(\frac{\partial S}{\partial t}\right)^2 + \left(1 - \frac{2M}{r}\right)\,
\left(\frac{\partial S}{\partial r}\right)^2 + m^2 = 0.
\end{equation}
We are interested in radial trajectories which are
independent of the angles $\theta , \varphi$.

For the definite energy state we obtain
\begin{equation}
- \frac{E^2}{\left(1 - \frac{2M}{r}\right)} + \left(1 -
\frac{2M}{r}\right)\, \left(\frac{dS_0}{dr}\right)^2 + m^2 =
0.\label{diff}
\end{equation}
Despite the fact that the Schwarzschild metric has two disjoint parts separated
by $r=2M$, we can nevertheless consider solutions of \eqref{diff} in these two regions and
glue them by going around the pole in the complex $r$-plane.
The solution is
\begin{equation} S_0 = \pm \int_0^{+\infty} \frac{dr}{\left(1 -
\frac{2M}{r}\right)}\, \sqrt{E^2 - m^2 \left(1 -
\frac{2M}{r}\right)},\label{intsch}
\end{equation}
where the limits of integration are chosen such that the particle goes through
the horizon $r=2M$. We focus on the integration through $r=2M$ since this is exactly
where the complex part of $S_0$ comes from. The $+ (-)$ sign in front of this integral indicates
that the particle is ingoing (outgoing). Although there is no classical path that
crosses $r=2M$ there are well established methods for dealing with
differential equations with singularities as in \eqref{diff}. The review article
by Brout and Spindel \cite{brout} or more recent work \cite{padman} \cite{padman1}
\cite{vagenas1} shows, in the present Schwarzschild case, how to deal with the 
paths which cross $r=2M$.

Because there is a pole at $r=2M$ along the path of
integration the integral will just be the Cauchy principle
value. The imaginary part of the principle value of \eqref{intsch}
is given by the contour integral over a
small half--loop going around the pole from below from
left to right. To explicitly take the imaginary part of the principle value we
make the change of variables $r-2M = \epsilon\, e^{{\rm i}\, \theta}$. Then
\begin{equation}
{\rm Im} S_0 = \pm \lim_{\epsilon \to 0} \int_{\pi}^{2\, \pi}
\frac{\left(2M + \epsilon\, e^{{\rm i}\, \theta}\right) \,
\epsilon\, e^{{\rm i}\, \theta} \, {\rm i}\, d\theta}{\epsilon \,
e^{{\rm i}\, \theta}}\, \sqrt{E^2 - m^2 \left(1 - \frac{2M}{2M +
\epsilon\, e^{{\rm i}\, \theta}}\right)} = \pm 2\,\pi\, M\,
E.\label{Hawef}
\end{equation}
All the above calculations can be easily performed for Reissner-Nordstrom
and Kerr black holes. All relevant formulas for the latter case can be
found in \cite{landau2}.

Using this result for ${\rm Im} S_0$ for the outgoing particle
the decay rate of the black hole is
$\Gamma \propto e^{\frac{-4\,\pi\,M\,E}{\hbar}}$.
This is just the Boltzmann weight with the temperature $T= \hbar /4\,\pi\,M$.
However, this disagrees with Hawking's value of the temperature by the factor
of 2. We now discuss this disagreement by examining different frames.
In turn we will work in the isotropic frame and the Painlev{\'e} frame.

If we make a generally covariant transformation of the
Schwarzschild coordinate frame \eqref{schwarz}
which involves only {\it spatial} coordinates this amounts to only a
change of integration variables in \eqref{intsch} or the variables
in \eqref{diff}. This does not change the result in \eqref{Hawef}.
For example, if we make a change of variables to isotropic coordinates as
in \cite{zerbini1} \cite{zerbini2}
\begin{equation}
\label{iso-coor}
r=\rho \left( 1 + \frac{M}{2 \rho} \right) ^2
\end{equation}
the Schwarzschild metric \eqref{schwarz} becomes
\begin{equation}
\label{iso-metric}
ds^2 = - \left( \frac{\rho - \frac{M}{2}}{\rho + \frac{M}{2}} \right)^2 dt^2
+\left( \frac{\rho + \frac{M}{2}}{\rho} \right)^4 \left( d \rho ^2 + \rho ^2 d \Omega ^2 \right).
\end{equation}
Now instead of \eqref{intsch} we find
\begin{equation} S_0 = \pm \int \frac{(\rho + \frac{M}{2})^3}{(\rho -\frac{M}{2}) \rho^2}
\sqrt{E^2 - m^2 \left( \frac{\rho -\frac{M}{2}}{\rho + \frac{M}{2}}\right)^2} ~ d \rho,\label{intiso}
\end{equation}
If one does the contour integration in the same manner as in \eqref{Hawef} by making a
semi-circular contour one apparently finds that Im $S_0 = \pm 4 \pi M E$. However one must also
deform the contour from \eqref{Hawef} using \eqref{iso-coor} and when this is done the semi-circular
contour of \eqref{Hawef} gets transformed into a quarter circle so that one gets $i \frac{\pi}{2} Residue$
rather than $i \pi Residue$. One could already guess this because from \eqref{iso-coor} $\rho \simeq
\sqrt{r}$ which for the contour in \eqref{Hawef} means the semi-circular contour becomes a
quarter circle. In detail
\begin{equation}
r=2M-\epsilon e ^{i \theta} = \rho + M +\frac{M^2}{4 \rho} \rightarrow
\rho = \frac{1}{2} \left( M + \epsilon e^{i \theta}
\pm  (2 M + \epsilon e^{i \theta} ) \sqrt{\epsilon} e^{i \theta /2} \right)
\end{equation}
The leading order in epsilon is now $\sqrt{\epsilon}$ so in the limit $\epsilon
\rightarrow 0$ we find from the above equation $\rho -\frac{M}{2} = M \sqrt{\epsilon}
e^{i \theta /2}$ instead of $r- 2M = \epsilon e^{i \theta}$. Thus one sees that the semi-circular
contour of the Schwarzschild frame gets transformed into a quarter circular in the isotropic
coordinate frame so that the result of integrating \eqref{intiso} is $i \frac{\pi}{2} Residue$ and
we find again Im$S_0 = \pm 2 \pi M E$. Note that in isotropic coordinates the spatial part
of the metric is no longer singular at the horizon. Thus both the Schwarzschild and isotropic
frame give the same temperature for the thermal radiation, but this temperature is
twice that given in the original quantum field theory inspired calculation \cite{hawking}.

The previous examples were related to one another via a transformation
of the spatial coordinates. Mathematically this just corresponds to a
change of variable between \eqref{intsch} and \eqref{intiso}, and this can not
change the results. However, one could consider a transformation which modifies the {\it time}
coordinate. For example, consider the transformation
\begin{equation}
dt' = dt + \frac{\sqrt{\frac{2M}{r}} \, dr}{1-\frac{2\,
M}{r}},\quad r' = r, \quad \Omega' = \Omega . \label{chan}
\end{equation}
With this the Schwarzschild metric takes the Painlev{\'e} form
\begin{equation}
ds^2 = - \left(1-\frac{2M}{r}\right)\, dt^2 +
2\sqrt{\frac{2M}{r}}\, dr\,dt + dr^2 + r^2 \, d\Omega^2  , \label{pmetric}
\end{equation}
where we have dropped the primes. The use of the Painlev{\'e} metric in the
tunneling picture was first considered in \cite{volovik} in the context
of condensed matter ``dumb" holes.
This metric is regular (i.e. does not have the horizon for the {\it incoming}
particles) at $r=2M$.
However the notion of time is changed with respect to the
Schwarzschild coordinates, so that the
Hamiltonian--Jacobi equation for the definite energy state becomes
\begin{equation}
- E^2 + \left(1- \frac{2\, M}{r}\right)\,
\left(\frac{dS_0}{dr}\right)^2 + 2\,\sqrt{\frac{2\, M}{r}}\,
E\,\frac{dS_0}{dr} + m^2 = 0.
\end{equation}
The solution of this equation is
\begin{equation} S_0 = - \int_{C}
\frac{dr}{1-\frac{2M}{r}}\,\sqrt{\frac{2M}{r}}\, E \pm \int_{C}
\frac{dr}{1-\frac{2M}{r}}\,\sqrt{E^2 - m^2 \left(1 -
\frac{2M}{r}\right)}.\label{ft}
\end{equation}
This result can not be obtained from \eqref{intsch} via a change of
integration variables because the transformation \eqref{chan} {\it does} affect
the time--like Killing vector. One can see that \eqref{ft} differs from \eqref{intsch}
by the first term. The first term in \eqref{ft}
arises from the coordinate change in \eqref{chan}, since
\begin{equation}
\int E dt + S_0 = \int E dt' - \int
\frac{dr}{1-\frac{2M}{r}}\,\sqrt{\frac{2M}{r}}\, E + S_0,
\end{equation}
where $S_0$ is given by \eqref{intsch}. Physically this
new coordinate system corresponds to $r$-dependent, singular shift of
the initial time. It can be shown
that the observer corresponding to the Schwarzschild metric detects the modes
which do not get converted into normalized modes in the metric \eqref{pmetric}
after the coordinate change \eqref{chan}.

If we choose the minus sign in \eqref{ft} and the contour $C$ as
before the result is
\begin{equation}
{\rm Im} S_0 = - 4\,\pi\, M\, E.\label{Haw1}
\end{equation}
This is twice the result of \eqref{Hawef} because the first integral in
\eqref{ft} gives the same contribution to the complex part of $S_0$
as the second one. We still see thermal
radiation in this reference frame but with a smaller temperature by
a factor of two: $T=\hbar / 8 \pi M$. This temperature agrees exactly
with Hawking's original result. In fact, Hawking in his
calculation did not use the Schwarzschild frame but rather used a
frame where the time $t'$ was related to the Schwarzschild time via
$dt' = dt + dr/(1 - \frac{2M}{r})$ which is similar to the transformation
which takes one from the Schwarzschild metric to the Painlev{\'e} metric.

We now explore this apparent discrepancy between the Hawking temperature
as calculated in the Schwarzschild and isotropic frames and the Painlev{\'e}
frame. The decay rate can in general be written as
\begin{equation}
\label{gamma}
\Gamma  =  e^{- 2 \frac{Im(S_0)}{\hbar}} = e^{- 2 \frac{Im \int p_r \, dr}{\hbar}} ~.
\end{equation}
In other words the integrands in \eqref{intsch} \eqref{intiso} and \eqref{ft}
are simply the momentum in the radial direction. It was pointed out in \cite{chowdhury}
that $\int p_r \, dr$ is not canonically invariant implying that the $\Gamma$
given in \eqref{gamma} is not canonically invariant and not a proper
observable. One can \cite{chowdhury} define a canonically invariant $\Gamma$
by using the canonically invariant quantity $\oint p_r \, dr$ i.e. one can
write the decay rate as
\begin{equation}
\label{gamma1}
\Gamma  = e^{- \frac{Im \oint p_r \, dr}{\hbar}} ~.
\end{equation}
Using \eqref{gamma1} for $\Gamma$ one can see that the Schwarzschild and
isotropic frames still give the same result as before, and now
the Painlev{\'e} frame also yields the same result. For the closed path
in the integral of \eqref{gamma1} consider a path which begins just outside
the horizon, $r_o$, crosses to just inside the horizon, $r_i$, and then returns
i.e.
\begin{equation}
\label{path}
\oint p_r \, dr = \int _{r_o} ^{r_i} p^+ _r dr + \int _{r_i} ^{r_o} p^- _r dr ~.
\end{equation}
where $p^+_r$ ($p^- _r$) are for the ingoing (outgoing) particles.
For Schwarzschild and isotropic frames $p^+ _r = -p^- _r$ so
taking into account the negative sign from reversing the order of
integration in \eqref{path} we find that for Schwarzschild and
isotropic metric  $\oint p_r \, dr = 2 \int p_r \, dr$. So for
these metrics $2 \int p_r \, dr$ is equivalent to the canonically
invariant $\oint p_r \, dr$. On the other hand for Painlev{\'e}
metric $p^+ _r =0$ while $p^- _r$ is twice the value of $p^- _r$
given by the Schwarzschild or isotropic metric, thus for the
Painlev{\'e} metric the entire contribution to $Im S_0$ comes from
the outgoing particle. The conclusion is that if one requires
$\Gamma$ be canonical invariant the tunneling calculation gives
the same answer for all three forms of the non-rotating black hole
metric. One finds thermal radiation, but with twice the
temperature found in the original Hawking calculation
\cite{hawking}. At this stage, without a more rigorous
calculation, we can conclude the following: either (i) the
tunneling calculations are not correct in detail as far as
calculating the temperature of the thermal radiation, or (ii) the
Hawking temperature for a non-rotating black hole really is twice
as large as given by the original calculations.

We should stress at this point that even in the original Hawking
derivation if one uses the Schwarzschild frame instead of the Kruskal
frame, one obtains a temperature twice that given in \cite{hawking}. 
As a final comment the factor of two difference between tunneling 
calculations and the original field-theoretic inspired calculation 
is not a new. In \cite{spindel} a factor of two difference was
found between the tunneling and field theoretic inspired
calculations of the Hawkings-Gibbons temperature in an expanding
de Sitter Universe. Additionally in \cite{thooft} it was argued
that when one combines the equivalence principle with quantum
mechanics that the Hawking temperature may in fact be twice that
of the original calculation. This factor of two difference for the
Schwarzschild black hole is investigated in more detail in
\cite{akhmedov}.

\section{Unruh radiation}

We now apply our procedure to obtain the thermal radiation
seen by a permanently accelerating reference system, i.e.
Unruh radiation \cite{unruh}. Starting from Minkowski space
we make a coordinate change to a permanently accelerating reference frame:
\begin{equation}
x_0 = \left(\frac{1}{a} + x\right)\, \sinh{(a\,t)}, \qquad x_1 =
\left(\frac{1}{a} + x\right)\, \cosh{(a\,t)} -
\frac{1}{a},\label{corch1}
\end{equation}
where $a$ is the acceleration. This gives the Rindler metric
\begin{equation}
ds^2 = - (1+a\,x)^2\, dt^2 + dx^2 + dy^2 +
dz^2.\label{Rind}
\end{equation}
The coordinate change \eqref{corch1} has introduced a horizon at $x=-1/a$.
The physical meaning of the horizon is as
follows: in a permanently accelerating system there are some
points behind the accelerating observer from which a classical, {\it free} particle
will never be able to catch up with the center of the accelerating reference system.

The Hamiltonian--Jacobi equation for this metric is
\begin{equation}
-\frac{1}{(1+a\,x)^2}\, \left(\frac{\partial S}{\partial t}\right)^2 +
\left(\frac{\partial S}{\partial \vec{x}}\right)^2 + m^2 =
0.\label{HJRin}
\end{equation}
One can also obtain this by applying the coordinate
transformation of \eqref{corch1} to the Hamiltonian--Jacobi
equation in the Minkowski space.
This coordinate change introduces a singularity at $x=-1/a$ in \eqref{HJRin}. As
a result the Hamiltonian--Jacobi equation for the constant
energy state in the Rindler space becomes
\begin{equation}
- \frac{E^2}{(1+ a\, x)^2} + \left(\frac{d S_0}{d x}\right)^2
+ m^2 = 0.
\end{equation}
For the rest of this section we consider solutions of the Hamiltonian--Jacobi equations
which depend only on one space coordinate: $x$. The solution of the Hamiltonian--Jacobi
equations in Minkowski space
does not have any imaginary part, while the one in the Rindler space
does. This can be seen directly from the formula:
\begin{equation}
{\rm Im} S_0 = \pm \int_C \frac{dx}{1 + a\,x}\, \sqrt{E^2 -
m^2\, (1 + a\, x)^2}.\label{Unef}
\end{equation}
As in the Schwarzschild case we must be careful with our
contour of integration since there is a pole at $x=-1/a$ on
the real axis. Also as in the Schwarzschild case we expect to get the
imaginary contribution to $S_0$ precisely from the integration around the pole.
The contour $C$ that we take is a
half--loop encircling the pole from below. As before the result is
given by the Cauchy principle value.
Explicitly $S_0  = \pm \frac{\pi \, {\rm i}\, E}{a}$.
The decay rate is $\Gamma \propto e^{-\frac{2\, \pi \, E}{\hbar a}},$ and
is canonically invariant since one can write $\Gamma = \exp (Im \oint p_r \, dr / \hbar)$,
as for the Schwarzschild and isotropic metrics.
This is the Boltzmann weight with a temperature equal to
$T=\frac{a \hbar}{2\, \pi}$ \cite{unruh}.

As in the previous case any coordinate changes which do not
affect the time will not effect the temperature, since
such transformations are simply changes of the integration variables in
\eqref{Unef}. Again the physical meaning of the effect
can be described as follows: classically a free particle moving from the
spatial infinity $x=-\infty$ to $x=+\infty$ can never catch up
with an observer who is moving with a constant acceleration in the same
direction. However, quantum mechanically the corresponding wave
can tunnel through the gravitational barrier. This effect
is described by the complex part of the classical action.

Thus, an observer who permanently accelerates (from $t=-\infty$ to
$t=+\infty$) with constant acceleration sees a thermal radiation
bath. This naturally leads one to ask if one sees thermal
radiation if the acceleration is for a finite time or if the
acceleration is due to rotation/orbiting. Using our procedure we
now address the rotating/orbiting system.

\section{Rotating and orbiting frames}

We now move on to the case of the rotating/orbiting reference frame.
Starting with the cylindrical coordinate system
$ds^2 = - dt^2 + dr^2 + r^2 \, d\varphi^2 + dz^2$
we transform to a permanently rotating reference frame via
$t=t',\, z=z',\, r=r',\, \varphi' = \varphi - \omega
\, t$ , where $\omega$ is the angular velocity. Then the metric becomes
\begin{equation}
ds^2 = - (1 - \omega^2 \, r^2)\, dt^2 + 2\, \omega \, r^2 \,
d\varphi \,dt + dr^2 + r^2 \, d\varphi^2 + dz^2.\label{rotm}
\end{equation}
This metric has a light radius at $\omega\, r = 1$. In the rotating/orbiting cases
the light radius plays a role similar to that of the horizons in the previous examples.
Physically this light radius comes from the fact that in a rigidly rotating frame at some
large enough radius points move with the velocity of light. The time
of this metric coincides with the time of the original Minkowski metric.

We now solve the Hamiltonian--Jacobi equation for the metric
\eqref{rotm}. The equation is
\begin{equation}
-\left(\frac{\partial S}{\partial t} - \omega \, \frac{\partial S}{\partial
\varphi}\right)^2 + \frac{1}{r^2}\, \left(\frac{\partial S}{\partial
\varphi}\right)^2 + \left(\frac{\partial S}{\partial r}\right)^2 + m^2 =
0. \label{HJrot}
\end{equation}
We are looking for solutions which are independent of $z$.
The action for the constant energy, $E$, and angular momentum,
$\mu$, state is $S = E\, t + \mu\, \varphi + S_0(r)$.
With this \eqref{HJrot} becomes
\begin{equation}
- \left(E - \omega\, \mu\right)^2 + \frac{\mu^2}{r^2} +
\left(\frac{d S_0}{dr}\right)^2 + m^2 = 0.
\end{equation}
We note that $\mu$ is the angular momentum both in the
Minkowski and in the rotating reference systems. However
the energy in the rotating frame is $E - \omega\,\mu$
while in the Minkowski frame it is $E$.

We are looking for the imaginary part of
\begin{equation}
S_0 = \pm \int_{r_2}^{r_1} \frac{dr}{r}\,
\sqrt{\left[(E-\omega\, \mu)^2 - m^2\right]\, r^2 -
\mu^2}.\label{Pol}
\end{equation}
The limits are $0< r_1 < 1/\omega < r_2$, so that the particle
trajectory crosses the light radius since this is were we expect
to get the imaginary contribution if there is one. The integrand
has the standard polar coordinate pole at $r=0$, but it does not have a
pole at, $r=1/\omega$, unlike the previous cases.

If we put $\omega = 0$, but keep $\mu$ finite and take the
limit of integration as $r_1 < \mu/\sqrt{E^2 - m^2}$,
then the integral in \eqref{Pol} has a
non-zero imaginary part. The physical meaning of the imaginary part is that a classical particle
with non--zero angular momentum can never reach the
center around which it rotates. Classically it can not go beyond the
limiting radius of $\mu/\sqrt{E^2 - m^2}$, but quantum mechanically the
corresponding wave can reach $r=0$. This latter effect, which shows itself
here through the imaginary part of the action,
has nothing to do with the radiation we are looking for.

Next we consider $\omega \neq 0$, and $\mu = 0$. Under these
conditions the integral in \eqref{Pol}
does not have an imaginary part. Although the metric
in \eqref{rotm} has a light radius, there is no obstacle for a
classical particle at the spatial infinity $r=+\infty$ to reach
the rotational center $r=0$, independently how fast it rotates.

The effect we are studying is supposed to come from the imaginary part due to
$\omega \, r = 1$. To see this clearly we make the following change
of the integration variables in \eqref{Pol}:
$r = 1/\omega + a$ so the part of the integral that crosses the light radius becomes
\begin{equation}
S_0 = \lim_{\epsilon \to 0} \int_{+\epsilon}^{-\epsilon}
\frac{da}{1 + a\, \omega} \, \sqrt{\left[(E - \omega\, \mu)^2 -
m^2\right]\, (1 + \omega \, a)^2 - (\omega\,\mu)^2}.
\end{equation}
Taking the integral and the limit gives $0$. For finite $\epsilon$
this integral can be complex, which simply means that $r =
1/\omega$ is in the classically forbidden zone ($1 / \omega <
\mu/\sqrt{E^2 - m^2}$), but the limit $\epsilon \rightarrow 0$ is
still zero. Thus, we see that quasi--classical tunneling
calculation tells us that there is no Hawking-like radiation for
the rotating system.

One may ask if an orbiting reference frame which is displaced from
$r=0$ by $R$ will see thermal radiation, i.e. consider a coordinate system
displaced from the origin by a radius $R$, and with this radius
orbiting around the origin with the angular velocity $\omega$.
The coordinate change from Minkowski space is
$\vec{r}' = \vec{r} - \vec{R}(t)$, where $\vec{r} =
(r,\varphi),$ and $\vec{R}(t) = (R, \omega\, t)$.
All other coordinates are left unchanged. As a result all axes
associated with the rotating particle stay always parallel to the ones of the
original, Minkowski reference system.

Starting from Minkowski metric, we obtain:
\begin{equation}
ds^2 = - (1 - R^2\,\omega^2) \, dt^2 - 2\, \omega \, R\,
\sin{(\omega\, t - \varphi)}\, dr\,dt +2\, \omega \, R \, r
\cos{(\omega\, t - \varphi)}\, d \varphi\, dt  +  dr^2 + r^2 \,d\varphi^2 +
dz^2. \label{one}
\end{equation}
This metric is time dependent. In order to apply to our method we need to
transform to a time independent metric. This is done by applying the transformation
$\varphi' = \varphi - \omega\, t$ on \eqref{one}. The result is
\begin{equation}
d s^2 = - \left[1 - \left(R^2 + r^2 +2 R r \cos \varphi \right)\,\omega^2\right]
\, dt^2 - 2\, \omega \, R\, \sin{\varphi}\, dr\,dt + 2 \, r \,
\omega \, ( R \cos \varphi + r) \, d\varphi\, dt + dr^2 + r^2 \,d\varphi^2 + dz^2.
\end{equation}
This metric is time independent and the axes are rigidly fixed to
the orbiting observer. The question of what kind of radiation, if
any, such a circularly moving observer will see was first studied
in \cite{letaw}. Recent work on the question of what kind of
radiation will be detected by an observer moving along some
general trajectory in Minkowski spacetime can be found in
\cite{c-unruh} \cite{c-unruh1}. The second reference in particular has a fairly
complete list of previous work. As with the rotating metric of
\eqref{rotm} this metric has a light radius at $1 - \left(R^2 +
r^2 +2 R r \cos \varphi \right)\,\omega^2 = 0$, but as for
\eqref{rotm} we can show that there is no imaginary contribution
to $S_0$, so again quasi-classical tunneling approach tells us
that there is no thermal radiation. According to the detailed
investigation performed  in \cite{AkhmSing} we can claim that
the quasi--classical tunneling approach gives correct results only
for backgrounds which have event horizons.

\section{Conclusions}

In this article we have given a general, quasi--classical
procedure based on the Hamilton-Jacobi equation for calculating
whether or not a particular gravitational background gives rise to
thermal radiation or not. This procedure works for backgrounds
which are stationary or have a time-like Killing vector. We have
not taken into account the back--reaction of the radiation on the
gravitational background. In order to take into account the
back--reaction one would have to solve coupled equations of
gravity and matter fields. 
We have shown that the quasi--classical tunneling calculations in
the Schwarzschild and isotropic frames give a temperature for the
thermal radiation which is twice as large as the original field
theoretical calculations. Furthermore, if one requires that the
decay rate, $\Gamma$, be canonically invariant then the
Painlev{\'e} frame also gives a temperature twice as large as the
original calculation. To decide between the quasi--classical
tunneling and field theoretic calculations one could try to
perform a non-perturbative Schwinger-like \cite{schwinger}
calculation for particle creation in a strong (gravitational)
field. In lieu of such a calculation we can conclude: either (i)
the tunneling calculation is wrong in detail or (ii) the Hawking
temperature of a non-rotating black hole really is twice that
given in the original calculation.

An interesting question is how to relate our quasi--classical
observations to the picture of a simple detector moving along a
particular trajectory \cite{unruh}, i.e. we would like to see how
the presence of an imaginary contribution to the action in the
tunneling calculation is related to the clicking rate of the
detector in question \cite{letaw} \cite{AkhmSing} \cite{c-unruh} \cite{c-unruh1}.

\begin{center}
\bf{Acknowledgments}
\end{center}

AET would like to thank A.Vainstein, M.Voloshin and especially
A.Morozov and J.Bjorken for illuminating discussions. This work
supported by the following grants: RFBR 04-02-16880 and the Grant from the President
of Russian Federation for support of scientific schools, and
a CSU Fresno International Activities Grant.

%%%%%%%%%%%%%%%%%%%%%%%%%%%%%%%%%%%%%%%%%%%

\end{document}